\documentclass[aps,prl,reprint,superscriptaddress]{revtex4-1}
\usepackage{blindtext}
\usepackage{graphicx}
\usepackage{amsmath}
\usepackage{multirow}
\usepackage{url}
\usepackage{color,soul}
\usepackage[colorlinks,
            linkcolor=blue,
            anchorcolor=blue,
            citecolor=blue,
            ]{hyperref}
\usepackage{setspace}

\begin{document}

\title{Topologically Protected Photonic Modes in Composite Quantum Hall/Quantum Spin Hall Waveguides}

\author{Shukai Ma}
\affiliation{Department of Physics, University of Maryland, College Park, Maryland 20742-4111, USA}
\author{Bo Xiao}
\affiliation{Department of Electrical and Computer Engineering, University of Maryland, College Park, Maryland 20742-3285, USA}
\author{Yang Yu}
\affiliation{School of Applied and Engineering Physics, Cornell University, Ithaca, New York 14853, USA}
\author{Kueifu Lai}
\affiliation{School of Applied and Engineering Physics, Cornell University, Ithaca, New York 14853, USA}
\author{Gennady Shvets}
\affiliation{School of Applied and Engineering Physics, Cornell University, Ithaca, New York 14853, USA}
\author{Steven M. Anlage}
\affiliation{Department of Physics, University of Maryland, College Park, Maryland 20742-4111, USA}
\affiliation{Department of Electrical and Computer Engineering, University of Maryland, College Park, Maryland 20742-3285, USA}

\begin{abstract}
Photonic topological systems, the electromagnetic analog of the topological materials in condensed matter physics, create many opportunities to design optical devices with novel properties. We present an experimental realization of the bi-anisotropic meta waveguide photonic system replicating both quantum Hall (QH) and quantum spin-Hall (QSH) topological insulating phases. With careful design, a composite QH-QSH photonic topological material is created and experimentally shown to support reflection-free edgemodes, a heterogeneous topological structure that is unprecedented in condensed matter physics. The effective spin degree of freedom of such topologically protected modes determines their unique pathways through these systems, free from backscattering and able to travel around sharp corners. {As an example of their novel properties, we experimentally demonstrate reflection-less photonic devices including a 2-port isolator, a unique 3-port topological device, and a full 4-port circulator based on composite QH and QSH structures}.
\end{abstract}

\maketitle

\section {Introduction}

During the past few years the realization of topological properties in photonic systems has paved the way for many novel applications and the creation of new states of light \cite{Haldane2008,Hasan2010,Kraus2012, Raghu2008, Xie2018, Zhang2017, Hou2018}. Translating concepts from condensed matter physics into the optical domain has come from mapping electronic Hamiltonians, such as the Kane-Mele Hamiltonian, into photonic platforms \cite{Kane2005,Raghu2008,Bernevig2006,Qi2011,Khanikaev2017}. The photonic topological insulators (PTIs) possess similar topological properties to their electronic counterparts and allow light traveling through sharp corners without reflection \cite{Khanikaev2017}. The observation of topologically protected surface waves (TPSWs) at microwave frequencies has been reported with biased magneto-optical photonic crystals with broken time-reversal invariance \cite{Wang2008, Wang2009, Skirlo2015,Longhi2015,Lu2016,Chen2017,Yang2019}. Due to the difficulty of achieving strong magnetic effects at optical frequencies, a class of time-reversal invariant PTIs has been proposed: these include coupled resonators \cite{Hafezi2011,Hafezi2013,Gao2015,Mittal2014,Leykam2018} and bi-anisotropic meta-waveguide (BMW) PTIs \cite{Khanikaev2012,Chen2014,Cheng2016,Lai2016,Slobozhanyuk2016,Xiao2016,Yang2018}. Recent studies revealed new designs of time-reversal-invariant PTIs based on emulating the quantum valley Hall (QVH) degree of freedom (DOF) in photonic systems \cite{Chen2018, Noh2018, Yang2018, Gao2017, GladsteinGladstone2018, Chen2018a}. Elegant Floquet topological photonic designs are also reported \cite{Fang2012a,Fang2012,Rechtsman2013,Rechtsman2013a}. In this manuscript, we focus on the experimental realization of quantum spin Hall (QSH) and quantum Hall (QH) PTIs in the context of the BMW structure. {We believe that this is the first-ever realization of a time-reversal invariance breaking heterogeneous topological system based on two different microscopic Hamiltonians, in either the photonic or electronic domains}.

The BMW approach to PTIs is unique in that it allows for integration of different photonic topological paradigms based on the same unperturbed structure. {Here we present the first demonstration of a non-reciprocal BMW structure and integrate it with a reciprocal PTI domain to create a completely new class of PTI materials}. In such composite BMW structures, a spin-momentum-locked TPSW can be spatially guided and filtered into desired channels at the junction points between different topological domains with high efficiency and free from backscattering \cite{Ma2017}. We begin with the simulation and experimental studies of the QH-PTIs followed by integration of QH-QSH PTIs. We are the first to observe TPSWs at the interface of the QH-QSH composite system which effectively serves as a 2-port PTI isolator. {We then introduce experimental demonstrations of more complex PTI structures, namely a unique topological Y-junction and a full 4-port circulator}. 
Our realized circulator structure has unique advantages over conventional photonic designs where the guided modes are not topologically protected. {The topologically trivial resonator-based circulators generally suffer from narrow bandwidth and are susceptible to backscattering due to impedance mismatches} \cite{Wang2005}. An alternative QH-PTI based circulator faces challenges with respect to device size (tens of wavelengths) and the lack of a spin DOF \cite{Qiu2011}. {On the contrary, the topological protection and spin DOF ensure the propagation of guided waves in heterogeneous BMW-PTI systems is backscatter-free, with higher directivity} \cite{Ma2017}. The utilization of topologically protected waves allows for a broad spectral operating range, compact design, integration with other PTI paradigms and potential for high-power operation.

\begin{figure}
\centering
\includegraphics[width=0.5\textwidth]{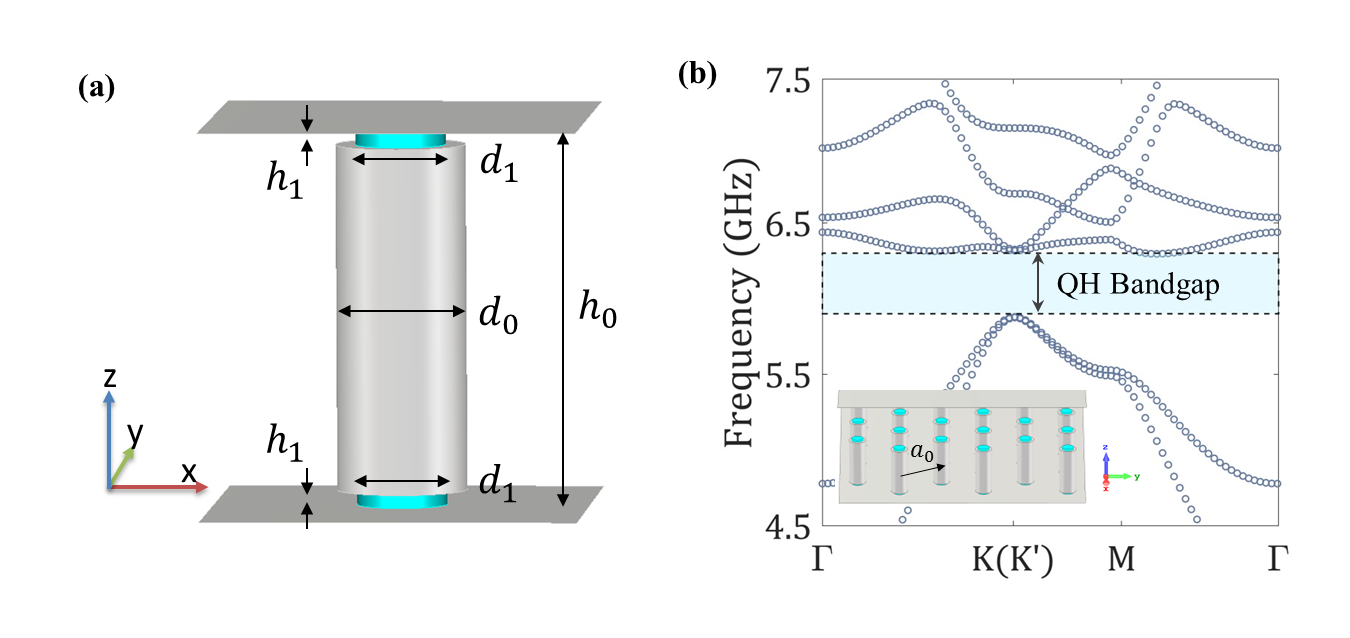}
\caption{\label{fig:QHbulk} (a) QH unit cell model with $h_0=36.8mm$, $h_1=1.45mm$, $d_0=12.7mm$ and $d_1=8.8mm$. A QH rod consists of two ferrite disks (blue) straddling a middle metallic rod, sandwiched by two metallic plates. (b) Photonic band structure for a QH bulk region. The horizontal-axis is the $k_x$ vector mapping an excursion in the 1st Brillouin zone and the vertical-axis is the eigenmode frequency. Under $H_z=1350 Oe$ the QH bulk bandgap (shaded area from 5.9 to 6.3 GHz) matches the QSH bandgap \cite{Xiao2016}. The inset offers an oblique view of a typical open-plate BMW hexagonal lattice with lattice constant $a_0=36.8mm$.}
\end{figure}

\section {QH-PTI System Design}

The BMW PTIs are based on an unperturbed structure of metallic rods arranged in a hexagonal lattice sandwiched between two metal plates \cite{Zandbergen2010,Bittner2010,Kuhl2010,Ma2015,Xiao2016,Ma2017}. The structure dimensions are carefully designed to ensure the TE and TM modes are degenerate at the Dirac points with the same group velocity in the photonic band structure (PBS). The two orbital DOF are emulated with left/right-hand circular polarizations and the two synthetic spin DOF by the in-phase and out-of-phase linear combinations of the TE and TM modes at the Dirac point \cite{Ma2017}. This mode crossing at the Dirac point can be destroyed by a broad range of spatial perturbations \cite{Ma2015,Cheng2016} thus creating a band gap, analogous to a bulk electronic insulator. In the present case, we create a QSH-PTI by adding an air gap between the rods and one plate, resulting in a bi-anisotropic response and an effective spin-orbit coupling (SOC) \cite{Xiao2016}. Alternatively one can create a QH-PTI by placing magneto-optical materials in the gap. For QSH and QH domains, the spin-Chern numbers $2C^{SOC}_{s,v}=\pm 1\times sgn(\Delta_{SOC})$ and a global Chern number $2C^T_{s,v}=sgn(\Delta_T)$, where $s=\uparrow , \downarrow$ is the spin state label and $v=K, K'$, are defined to quantify the topological properties of the upper and lower bands \cite{Ma2017}. The spin-orbit coupling $\Delta_{SOC}$ and time-reversal invariance breaking $\Delta_T$ perturbations are defined as the overlap integrals of unperturbed modes' field components within the perturbed volume in the QSH and QH regions, respectively. An interface separating two PTIs with different Chern numbers will support edge states that are spectrally localized in the bulk bandgap \cite{Ma2015,Khanikaev2012,Haldane2008}. Both QSH and QVH types of BMW have been experimentally realized \cite{Xiao2016, Yang2018,Gao2017}, leaving the experimental demonstration of the QH-BMW structure un-reported. Recent theoretical studies \cite{Ma2017} pave the way for realizing QH-PTIs in a BMW base by sandwiching biased gyromagnetic materials on top and bottom of the metallic rod (Fig. \ref{fig:QHbulk}(a)). {In contrast with} \cite{Ma2017} {which characterized the ferrites using a dielectric permittivity $\epsilon_r=1$ and a simplified permeability tensor, here the QH region is re-designed in order to incorporate realistic ferrite properties}. Figure. \ref{fig:QHbulk} (a) and (b) show the bulk QH unit-cell model and its PBS, respectively. This band diagram is obtained with first-principles electromagnetic simulation using COMSOL Multiphysics. The ferrites are characterized by $\epsilon_r=14$ and the Polder permeability tensor \cite{Polder1949} (see SM \cite{supmat}). As shown in Fig. \ref{fig:QHbulk}(b), under an external bias magnetic field of $H_z=1350 Oe$, a bandgap opens throughout the first Brillouin Zone, and matches closely with that of our existing QSH bulk structure \cite{Ma2015,Xiao2016}. {The degeneracy of TE and TM modes in the vicinity of the Dirac point is well-preserved as shown in Fig.} \ref{fig:QHbulk} {(b)}. 

\begin{figure}
\centering
\includegraphics[width=0.45\textwidth]{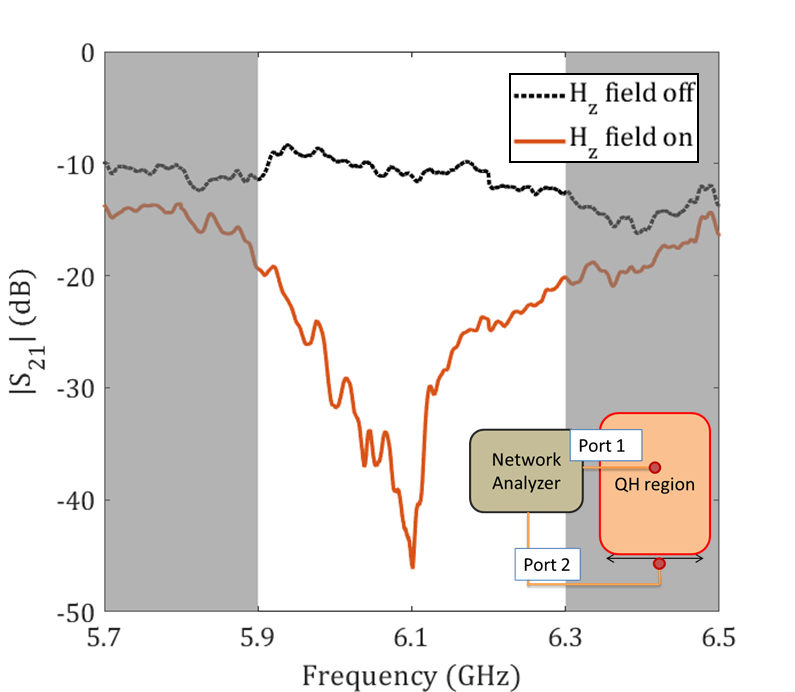}
\caption{\label{fig:QHS21a} $|S_{21}|$ transmission results through a bulk QH structure in the $\Gamma-K(K')$ direction for both magnetized and un-magnetized ferrites. With $H_z$ on (magnetized ferrites, red curve), we observe a decrease in transmission S-parameter at 5.9-6.3 GHz, which is caused by the absence of bulk modes inside the bandgap region. Inset: schematic of the experiment (see SM \cite{supmat} for more details).}
\end{figure}

We next conduct transmission measurements through the proposed QH bulk structure. {The gyromagnetic disks are made of commercial Yttrium Garnet ferrites which are magnetized along the $z$ axis using Nd-Fe-B permanent magnets located outside the structure on the top and bottom plates (see SM \cite{supmat})}. The QH region contains a $14 \times 5$ ferrite-loaded lattice. Experimental results are presented as transmission S-parameters in Fig. \ref{fig:QHS21a}. The inset of Fig. \ref{fig:QHS21a} shows the measurement method: the two ports of the Vector Network Analyzer (VNA) are connected to antennas at the center of a QH bulk structure and its perimeter. By switching on the H-field to magnetize the ferrites, we observed an averaged 20 dB decrease of transmission from port 1 to port 2 from 5.95 to 6.2 GHz which is slightly narrower compared to the photonic bandgap predicted in Fig. \ref{fig:QHbulk}(b).

\section {Composite QH-QSH Systems}

We next construct a heterogeneous PTI structure with two topological phases: QH and QSH PTIs. This composite PTI structure has an interface (Fig. S3) that will support topologically protected edgemodes (Fig. S2) in the operating bandwidth (the shaded area in Fig. \ref{fig:QHbulk}(b)) inside the original bandgap frequencies for both QSH and QH bulk structures. The key requirement is to maintain the spin degeneracy of the modes in the composite structure, which allows edgemodes propagating without reflection \cite{Ma2017}. Simulation results for a QH-QSH supercell structure (Fig. S2) show that in contrast to a QSH-QSH interface \cite{Xiao2016} where there are two edgemodes with opposite group velocities, a QH-QSH interface will only support one-way spin-locked TPSWs at the $K(K')$ points. 

\begin{figure}
\centering
\includegraphics[width=0.5\textwidth]{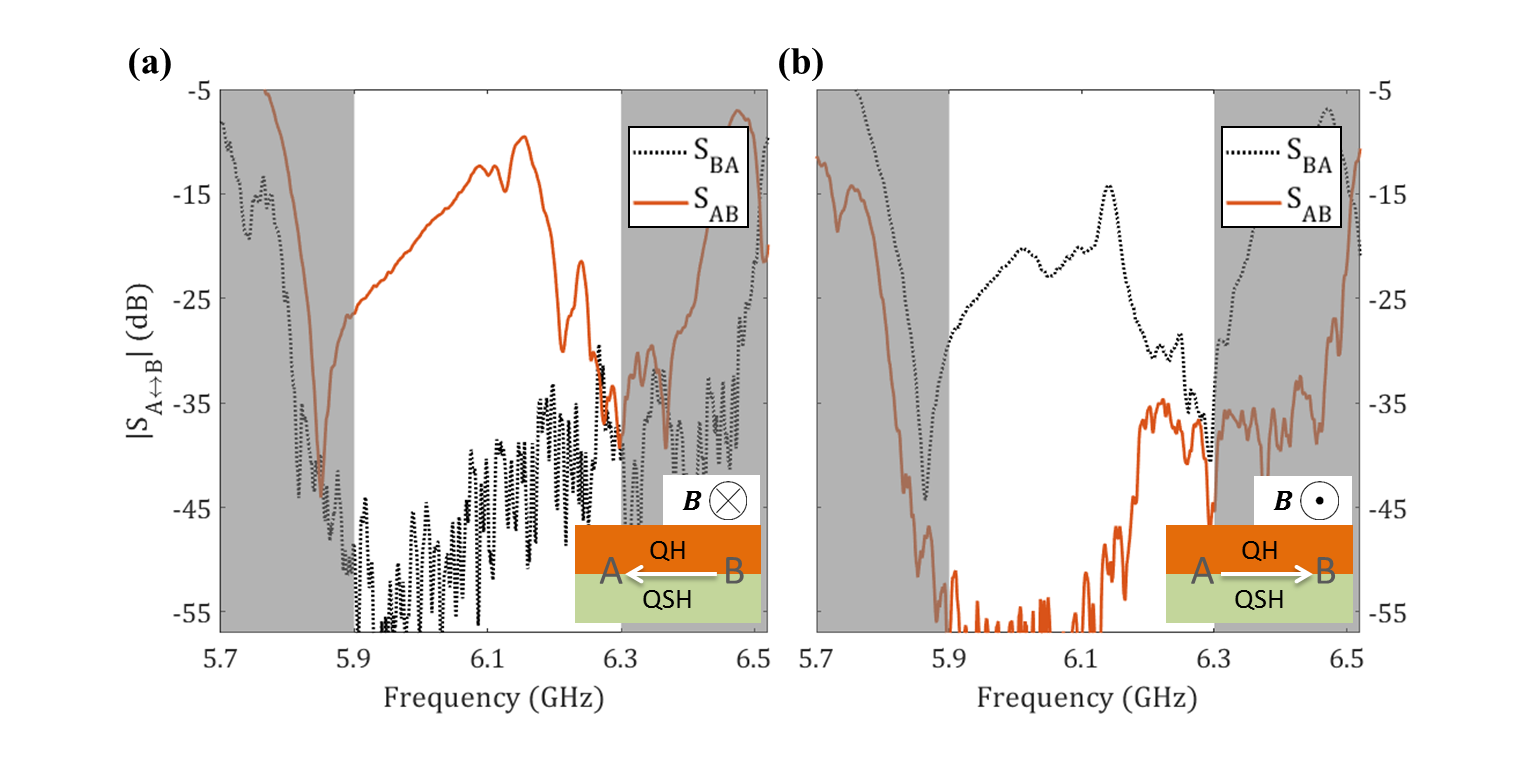}
\caption{\label{fig:QSHQHexp} (a) and (b) are the measured edgemode transmission along the QH-QSH interface with opposite H-field directions applied to the QH region. The black-dot(red-line) curves represent the right(left)-going waves. The insets show the locations of port A and B along with the direction of the applied H-field. The white arrows mark the expected edgemode propagating directions. {We find that the edgemode propagation direction is reversed when the biasing magnetic field direction is inverted}. }
\end{figure}

The composite QH-QSH PTI structure has $39 \times 5$ BMW rods in both regions (Fig. S3). We use the VNA to conduct 2-port measurements for edgemode transmission along the QH-QSH interface. {In Fig.} \ref{fig:QSHQHexp} {(a)/(b), the left/right-going transmission $S_{AB}/S_{BA}$ for the two possible magnetic field biasing orientations is 35 dB higher than the transmission in the other direction in the joint band gap frequency range}. These experimental results show that a QH-QSH interface can only support an edgemode with one synthetic spin direction. 
The backscattering-free property of an edgemode is also examined in Fig. S4. The inversion of applied H-field in the z-direction will invert the sign of the normalized bandgap $\Delta_T$ for QH-PTIs \cite{Ma2017}. With the QSH-PTIs unchanged, the supported edgemode at the QH-QSH interface will flip its spin, and more importantly invert its spin-locked propagation direction. {The topologically protected edgemodes ensure strong suppression of the reflected flow of waves in the un-desired direction}. The QH-QSH structure serves as an effective 2-port PTI isolator which can be turned on/off by external H-field (Fig. S7). The isolator's allowed propagating direction can be switched in real-time by the inversion of the H-field.

\begin{figure}
\centering
\includegraphics[width=0.5\textwidth]{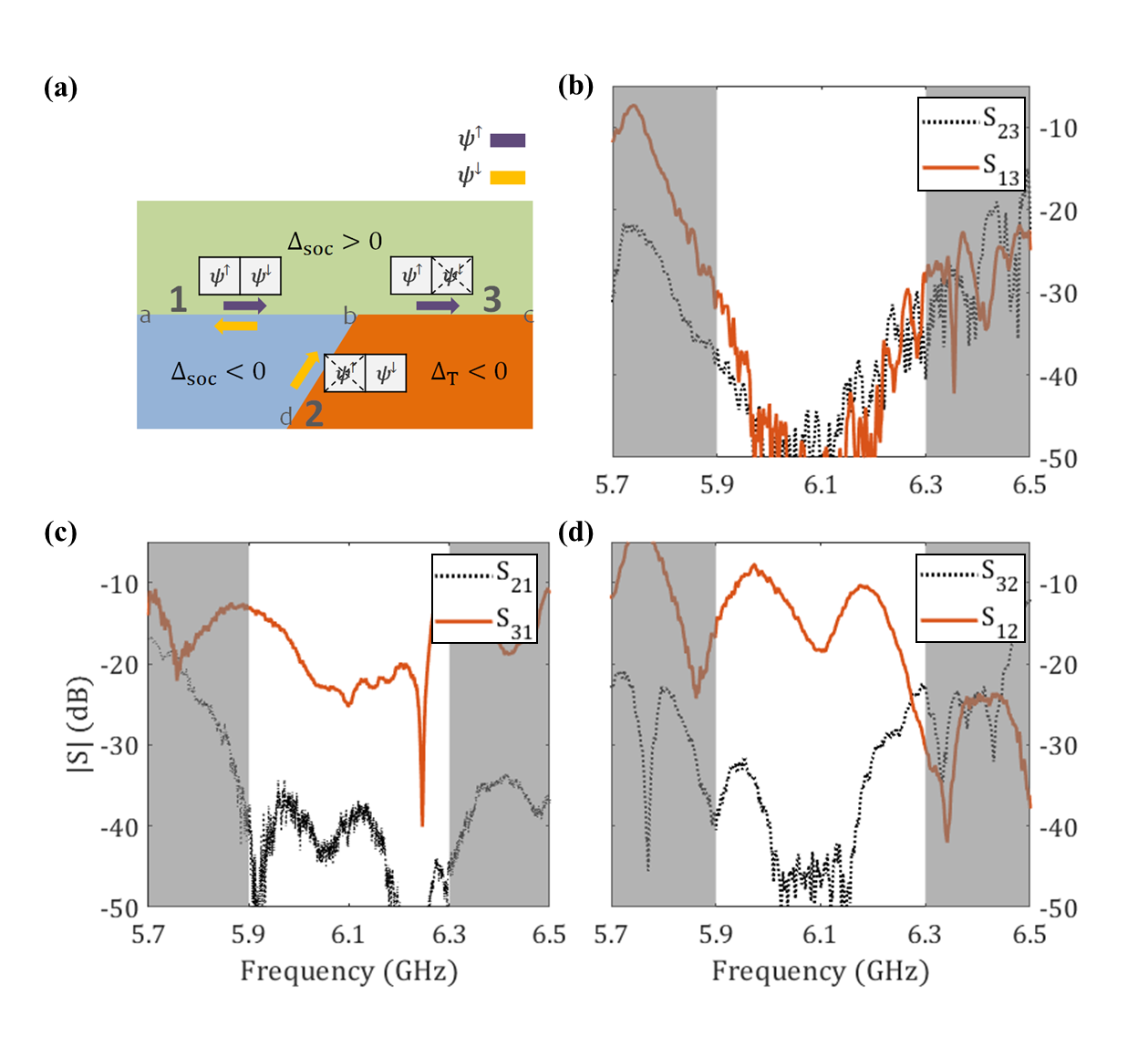}
\caption{\label{fig:Yexp} (a): The Y-junction design consist of three different topological domains: two QSH domains (green and blue) and a QH region (red), creating three interfaces: ab, bd and bc. The $\psi^{\uparrow/\downarrow}$ represents an edgemode with spin up/down. Purple/yellow arrows: propagation directions of $\psi^{\uparrow/\downarrow}$  modes. Plot (b-d) are the transmission measurements of the Y-junction with $-H_z$ in the QH region. The edgemode will follow the $2 \rightarrow 1$ and the $1 \rightarrow 3$ paths strictly without backscattering. Port 3 becomes a forbidden source as shown in (b).}
\end{figure}

\section {Topological Y-junction}

With the experimental realizations of both QSH-QSH \cite{Xiao2016} and QH-QSH waveguides, we next propose a `Y-junction' PTI structure which mixes three different topological phases (shown in Fig. S8 and schematically in Fig. \ref{fig:Yexp}(a)). {The 3-port Y-junction possesses unique functions compared to more traditional devices, such as a 3-port circulator}. With its spin DOF and propagation direction locked together, the edgemode transmission behavior inside the Y-junction can be predicted theoretically. For instance, an edgemode with either spin up ($\psi^{\uparrow}$) or down ($\psi^{\downarrow}$) can exist at interface `ab', with their propagating direction restricted to right and left-going, respectively. Governed by the direction of the external H-field in the QH-region, a $-H_z$ field will dictate that the Y-junction's port 3 (2) becomes a forbidden source (receiver). To illustrate a forbidden source (receiver) experimentally, note the decrease of both $S_{23}$ and $S_{13}$ ($S_{23}$ and $S_{21}$) from 5.9-6.3 GHz shown in Fig. \ref{fig:Yexp}(b) (Fig. \ref{fig:Yexp}(b) and (c)). Because the interface `bc' can only support an up-spin mode $\psi^{\uparrow}$, the signal injected at port 3 can not travel left to port 1 or 2. In Fig. \ref{fig:Yexp}(c) the amplitude of $S_{31}$ is 20 dB higher than $S_{21}$, indicating that a right-going wave transmitted from port 1 will travel to port 3 instead of port 2. This choice of path is dictated by the fact that such an up-spin TPSW can only propagate along interface `bc' but not `bd'. For similar reasons, a down-spin wave $\psi^{\downarrow}$  transmitted by port 2 will only go to port 1 as shown in Fig. \ref{fig:Yexp}(d). 
{The ideal S-matrix of the topological Y-junction can be written as} $S_{Topo-Y}=\begin{bmatrix} 0 & 1 & 0 \\ 0 & 0 & 0 \\ 1 & 0 & 0 \end{bmatrix}$, {which differs from a traditional 3-port circulator} $S_{Circ}=\begin{bmatrix} 0 & 1 & 0 \\ 0 & 0 & 1 \\ 1 & 0 & 0 \end{bmatrix}$ {which creates a} $1\rightarrow3\rightarrow2\rightarrow1$ {circulation}. The functionality of a topological Y-junction is similar to the active quasi-circulator \cite{Mung2019}.
Inversion of the H-field direction will turn port 3 (2) into a forbidden receiver (source), due to the change in propagating edgemode spin polarization at the two QH-QSH interfaces (observed but not shown here) \cite{Ma2017}. 
The topological Y-junction will act as a building block for more sophisticated structures, such as the 4-port circulator. {The demonstrated Y-junction can serve as a versatile platform for many novel photonic applications based on various combinations of topological domains. For example, one can construct a backscattering-free photonic combiner with a QSH region and two QH regions with opposite H-field directions that leads edgemodes from two paths to merge onto the same third path with no backscattering. Furthermore, an inversion of biasing H-fields in the QH domains will turn the photonic combiner into a spin filter that divides photons with different synthetic spins into different paths, effectively creating a Stern-Gerlach device for the spin DOF of light. Such junctions may find application in quantum communication, Boolean networks models, and photonic devices based on manipulating photons with different polarization states}.

\section {Topological 4-port Circulator}

\begin{figure}
\centering
\includegraphics[width=0.5\textwidth]{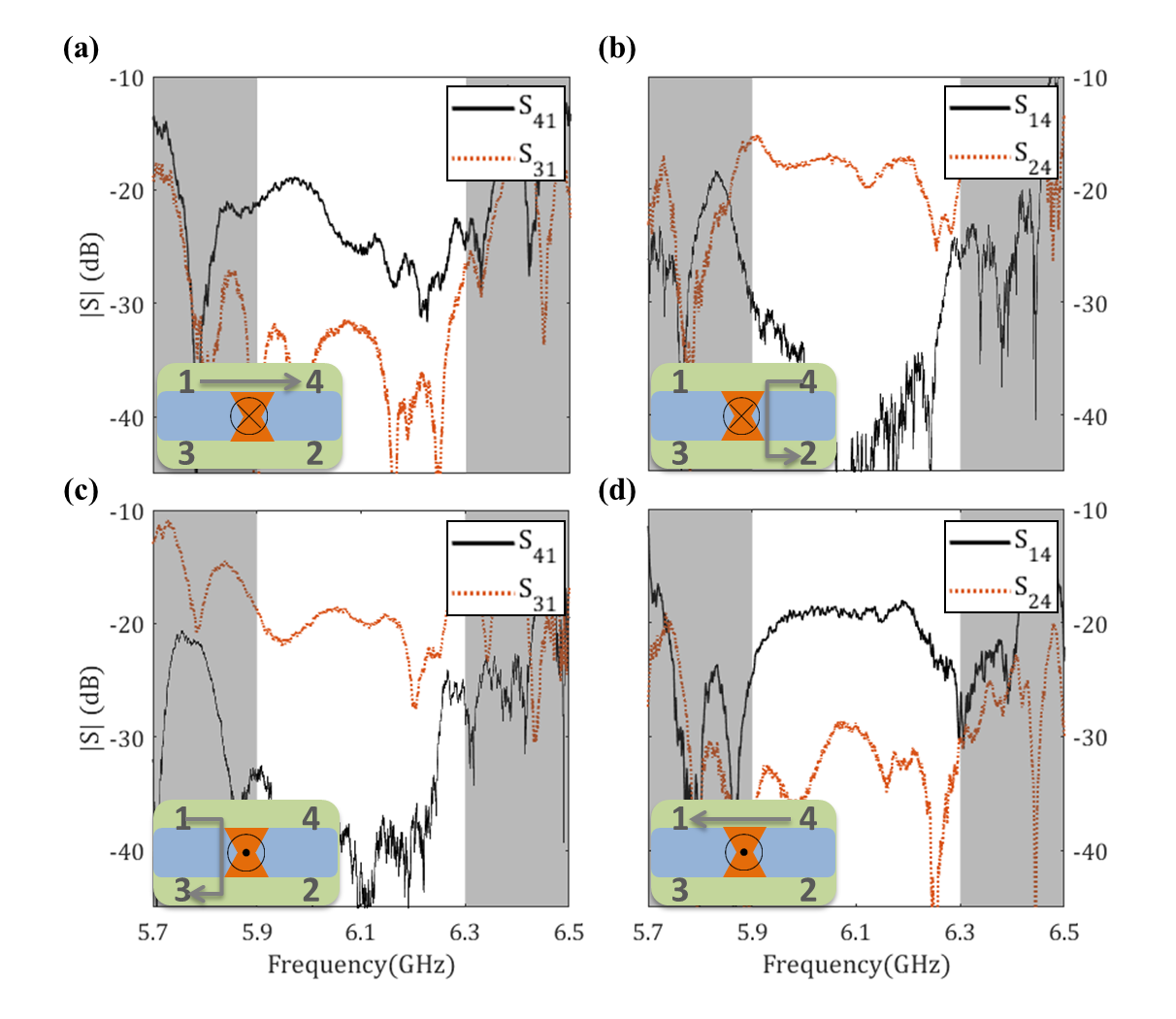}
\caption{\label{fig:circ2} Experimental results of edge waves propagating in a topologically protected 4-port circulator. The QH (QSH with $\Delta_{SOC}>0$ and $\Delta_{SOC}<0$) domains are color-coded as red (green and blue) for all insets. The applied H-field direction is inverted between plots (a-b) and (c-d), creating a clockwise and counter-clockwise circulation around the center QH island, respectively.}
\end{figure}

Novel non-reciprocal structure designs, like isolators and circulators, have been proposed recently due to their important roles in photonic and microwave circuits \cite{Seif2018, Fang2017, Shen2018,Ni2018, Jia2019, Fleury2014, Maayani2018, Khanikaev2015}. {After the experimental demonstration of the Y-junction structure, we are in position to integrate different PTIs into another practical structure, namely a 4-port circulator}. In contrast to realizing optical circulators with only 2D gyromagnetic PTIs \cite{Wang2005}, here we report the experimental realization of the combined PTI system with different types of topological phases \cite{Ma2017,Khanikaev2017}. Similar to the design in \cite{Ma2017}, the realized 4-port BMW circulator consists of a center QH island and four surrounding QSH regions with alternating $\Delta_{SOC}$ signs (Fig. \ref{fig:circ2}). The directions of both QSH-QSH and QH-QSH waveguides are chosen to be in the $K(K')$ directions for optimized edgemode propagation.

The operation of the BMW circulator is as follows. To start with, consider the trajectory of the up-spin edgemode $\psi^{\uparrow}$ which is launched from port 1 and propagating right along the QSH(green)-QSH(blue) waveguide (inset of Fig. \ref{fig:circ2}(a)). {When the edgemode arrives at the QSH-QH-QSH Y-junction, there exist two QH-QSH waveguides (`red-blue' and `red-green') which support edgemodes with opposite spins}. {The topological nature of these modes will lead the wave to only one of the two QH-QSH waveguides, without backscattering} \cite{Ma2017}: at the first Y-junction, the spin up wave $\psi^{\uparrow}$ will merge on to the horizontal QH-QSH interface; at the second Y-junction, it will choose the QSH-QSH interface over the vertical QH-QSH interface because that QH-QSH interface does not support spin up edgemodes. As shown in Fig. \ref{fig:circ2}(a), the measured transmission from port 1 to 4 ($S_{41}$) enjoys an averaged 15 dB boost compared to the competing vertical path (port 1 to 3) in the bulk band gap from 5.9 to 6.3 GHz. Combined with the other three guided edgemode paths, a clear circulation pattern is observed around the center QH-island.  

The inversion of the $H_z$ direction will change the sign of $\Delta_T$ of the QH region and further lead to the flip of spins of the propagating modes at all QH-QSH interfaces. Considering again the same up-spin TPSW launched from port 1: instead of choosing the horizontal QH-QSH interface, it will merge on to the vertical QH-QSH interface and finally flow to port 3. As shown in Fig. \ref{fig:circ2}(c), $S_{31}$ (the transmission from port 1 to port 3) is now higher than $S_{41}$. To conclude: we observed the change of circulation direction from counter-clockwise (Fig. \ref{fig:circ2} (a) and (b)) to clockwise (Fig. \ref{fig:circ2} (c) and (d)) controlled by {the external H-field biasing direction} on the QH BMW region. The isolation of the circulator design is 20 dB on average, the input power is +5 dBm limited only by the VNA, and the frequency range is 5.9-6.3 GHz which can be tuned by scaling the structural dimensions.

{Note that the parallel QSH-QSH waveguides can have cross-coupling issues caused by the finite-size limitation of the structure (see SM \cite{supmat})}. {The devices also face energy dissipation due to the finite line-width of the ferrite, and enhanced insertion loss due to imperfect coupling between the antenna and guided modes in the structure}. Implementation of a rotating dipole source will enhance the power delivery by directly and efficiently exciting uni-directional edgemodes on the QSH-QSH interfaces \cite{Xiao2016}.


{In conclusion, we experimentally realized the first BMW-type bulk QH-PTI materials and created a bandgap that matches the QSH-PTI structure} \cite{Xiao2016} {while maintaining spin degeneracy}. We then observed the appearance of a QH bandgap experimentally by applying external magnetic field to a ferrite-loaded BMW. We are the first to create a QH-QSH interface system and observe spin-momentum-locked guided waves whose propagation depends on the pseudo-spin of the excitation as well as the direction of the magnetization of the QH region. {Based on this result, we proposed and realized several new reflection-less photonic devices, such as a 2-port isolator and a 3-port topological Y-junction, and observed guided edgemodes that followed spin-dependent paths expected from theoretical predictions}. We constructed a 4-port BMW circulator and observed clear circulation of edgemodes, where the circulation direction is dictated by the magnetization direction of the ferrites. The ability to construct composite photonic systems with both QSH and QH phases offers more flexible ways to study new phenomena that are difficult to achieve in condensed matter systems. 

\begin{acknowledgments}
This work was supported by ONR under Grant Nos. N000141512134 and N000141912481, AFOSR COE Grant FA9550-15-1-0171, DOE under grant DESC 0018788, the National Science Foundation (NSF) under Grants No. DMR-1120923, No. PHY-1415547 and No. ECCS-1158644.
\end{acknowledgments}


\begin{thebibliography}{57}%
\makeatletter
\providecommand \@ifxundefined [1]{%
 \@ifx{#1\undefined}
}%
\providecommand \@ifnum [1]{%
 \ifnum #1\expandafter \@firstoftwo
 \else \expandafter \@secondoftwo
 \fi
}%
\providecommand \@ifx [1]{%
 \ifx #1\expandafter \@firstoftwo
 \else \expandafter \@secondoftwo
 \fi
}%
\providecommand \natexlab [1]{#1}%
\providecommand \enquote  [1]{``#1''}%
\providecommand \bibnamefont  [1]{#1}%
\providecommand \bibfnamefont [1]{#1}%
\providecommand \citenamefont [1]{#1}%
\providecommand \href@noop [0]{\@secondoftwo}%
\providecommand \href [0]{\begingroup \@sanitize@url \@href}%
\providecommand \@href[1]{\@@startlink{#1}\@@href}%
\providecommand \@@href[1]{\endgroup#1\@@endlink}%
\providecommand \@sanitize@url [0]{\catcode `\\12\catcode `\$12\catcode
  `\&12\catcode `\#12\catcode `\^12\catcode `\_12\catcode `\%12\relax}%
\providecommand \@@startlink[1]{}%
\providecommand \@@endlink[0]{}%
\providecommand \url  [0]{\begingroup\@sanitize@url \@url }%
\providecommand \@url [1]{\endgroup\@href {#1}{\urlprefix }}%
\providecommand \urlprefix  [0]{URL }%
\providecommand \Eprint [0]{\href }%
\providecommand \doibase [0]{http://dx.doi.org/}%
\providecommand \selectlanguage [0]{\@gobble}%
\providecommand \bibinfo  [0]{\@secondoftwo}%
\providecommand \bibfield  [0]{\@secondoftwo}%
\providecommand \translation [1]{[#1]}%
\providecommand \BibitemOpen [0]{}%
\providecommand \bibitemStop [0]{}%
\providecommand \bibitemNoStop [0]{.\EOS\space}%
\providecommand \EOS [0]{\spacefactor3000\relax}%
\providecommand \BibitemShut  [1]{\csname bibitem#1\endcsname}%
\let\auto@bib@innerbib\@empty
\bibitem [{\citenamefont {Haldane}\ and\ \citenamefont
  {Raghu}(2008)}]{Haldane2008}%
  \BibitemOpen
  \bibfield  {author} {\bibinfo {author} {\bibfnamefont {F.~D.~M.}\
  \bibnamefont {Haldane}}\ and\ \bibinfo {author} {\bibfnamefont
  {S.}~\bibnamefont {Raghu}},\ }\href {\doibase 10.1103/PhysRevLett.100.013904}
  {\bibfield  {journal} {\bibinfo  {journal} {Physical Review Letters}\
  }\textbf {\bibinfo {volume} {100}},\ \bibinfo {pages} {013904} (\bibinfo
  {year} {2008})}\BibitemShut {NoStop}%
\bibitem [{\citenamefont {Hasan}\ and\ \citenamefont {Kane}(2010)}]{Hasan2010}%
  \BibitemOpen
  \bibfield  {author} {\bibinfo {author} {\bibfnamefont {M.~Z.}\ \bibnamefont
  {Hasan}}\ and\ \bibinfo {author} {\bibfnamefont {C.~L.}\ \bibnamefont
  {Kane}},\ }\href {\doibase 10.1103/RevModPhys.82.3045} {\bibfield  {journal}
  {\bibinfo  {journal} {Reviews of Modern Physics}\ }\textbf {\bibinfo {volume}
  {82}},\ \bibinfo {pages} {3045} (\bibinfo {year} {2010})}\BibitemShut
  {NoStop}%
\bibitem [{\citenamefont {Kraus}\ \emph {et~al.}(2012)\citenamefont {Kraus},
  \citenamefont {Lahini}, \citenamefont {Ringel}, \citenamefont {Verbin},\ and\
  \citenamefont {Zilberberg}}]{Kraus2012}%
  \BibitemOpen
  \bibfield  {author} {\bibinfo {author} {\bibfnamefont {Y.~E.}\ \bibnamefont
  {Kraus}}, \bibinfo {author} {\bibfnamefont {Y.}~\bibnamefont {Lahini}},
  \bibinfo {author} {\bibfnamefont {Z.}~\bibnamefont {Ringel}}, \bibinfo
  {author} {\bibfnamefont {M.}~\bibnamefont {Verbin}}, \ and\ \bibinfo {author}
  {\bibfnamefont {O.}~\bibnamefont {Zilberberg}},\ }\href {\doibase
  10.1103/PhysRevLett.109.106402} {\bibfield  {journal} {\bibinfo  {journal}
  {Physical Review Letters}\ }\textbf {\bibinfo {volume} {109}},\ \bibinfo
  {pages} {106402} (\bibinfo {year} {2012})}\BibitemShut {NoStop}%
\bibitem [{\citenamefont {Raghu}\ and\ \citenamefont
  {Haldane}(2008)}]{Raghu2008}%
  \BibitemOpen
  \bibfield  {author} {\bibinfo {author} {\bibfnamefont {S.}~\bibnamefont
  {Raghu}}\ and\ \bibinfo {author} {\bibfnamefont {F.~D.~M.}\ \bibnamefont
  {Haldane}},\ }\href {\doibase 10.1103/PhysRevA.78.033834} {\bibfield
  {journal} {\bibinfo  {journal} {Physical Review A}\ }\textbf {\bibinfo
  {volume} {78}},\ \bibinfo {pages} {033834} (\bibinfo {year}
  {2008})}\BibitemShut {NoStop}%
\bibitem [{\citenamefont {Xie}\ \emph {et~al.}(2018)\citenamefont {Xie},
  \citenamefont {Wang}, \citenamefont {Zhu}, \citenamefont {Lu}, \citenamefont
  {Wang},\ and\ \citenamefont {Chen}}]{Xie2018}%
  \BibitemOpen
  \bibfield  {author} {\bibinfo {author} {\bibfnamefont {B.-y.}\ \bibnamefont
  {Xie}}, \bibinfo {author} {\bibfnamefont {H.-F.}\ \bibnamefont {Wang}},
  \bibinfo {author} {\bibfnamefont {X.-y.}\ \bibnamefont {Zhu}}, \bibinfo
  {author} {\bibfnamefont {M.-H.}\ \bibnamefont {Lu}}, \bibinfo {author}
  {\bibfnamefont {Z.~D.}\ \bibnamefont {Wang}}, \ and\ \bibinfo {author}
  {\bibfnamefont {Y.-f.}\ \bibnamefont {Chen}},\ }\href {\doibase
  10.1364/OE.26.024531} {\bibfield  {journal} {\bibinfo  {journal} {Optics
  Express}\ }\textbf {\bibinfo {volume} {26}},\ \bibinfo {pages} {24531}
  (\bibinfo {year} {2018})}\BibitemShut {NoStop}%
\bibitem [{\citenamefont {Zhang}(2017)}]{Zhang2017}%
  \BibitemOpen
  \bibfield  {author} {\bibinfo {author} {\bibfnamefont {F.}~\bibnamefont
  {Zhang}},\ }\href@noop {} {\bibfield  {journal} {\bibinfo  {journal}
  {Science}\ }\textbf {\bibinfo {volume} {358}},\ \bibinfo {pages} {1075}
  (\bibinfo {year} {2017})}\BibitemShut {NoStop}%
\bibitem [{\citenamefont {Hou}\ \emph {et~al.}(2018)\citenamefont {Hou},
  \citenamefont {Li}, \citenamefont {Luo}, \citenamefont {Gu},\ and\
  \citenamefont {Zhang}}]{Hou2018}%
  \BibitemOpen
  \bibfield  {author} {\bibinfo {author} {\bibfnamefont {J.}~\bibnamefont
  {Hou}}, \bibinfo {author} {\bibfnamefont {Z.}~\bibnamefont {Li}}, \bibinfo
  {author} {\bibfnamefont {X.-W.}\ \bibnamefont {Luo}}, \bibinfo {author}
  {\bibfnamefont {Q.}~\bibnamefont {Gu}}, \ and\ \bibinfo {author}
  {\bibfnamefont {C.}~\bibnamefont {Zhang}},\ }\href
  {https://arxiv.org/pdf/1808.06972.pdf http://arxiv.org/abs/1808.06972} {\
  (\bibinfo {year} {2018})},\ \Eprint {http://arxiv.org/abs/1808.06972}
  {arXiv:1808.06972} \BibitemShut {NoStop}%
\bibitem [{\citenamefont {Kane}\ and\ \citenamefont {Mele}(2005)}]{Kane2005}%
  \BibitemOpen
  \bibfield  {author} {\bibinfo {author} {\bibfnamefont {C.~L.}\ \bibnamefont
  {Kane}}\ and\ \bibinfo {author} {\bibfnamefont {E.~J.}\ \bibnamefont
  {Mele}},\ }\href {\doibase 10.1103/PhysRevLett.95.226801} {\bibfield
  {journal} {\bibinfo  {journal} {Physical Review Letters}\ }\textbf {\bibinfo
  {volume} {95}},\ \bibinfo {pages} {226801} (\bibinfo {year}
  {2005})}\BibitemShut {NoStop}%
\bibitem [{\citenamefont {Bernevig}\ and\ \citenamefont
  {Zhang}(2006)}]{Bernevig2006}%
  \BibitemOpen
  \bibfield  {author} {\bibinfo {author} {\bibfnamefont {B.~A.}\ \bibnamefont
  {Bernevig}}\ and\ \bibinfo {author} {\bibfnamefont {S.-C.}\ \bibnamefont
  {Zhang}},\ }\href {\doibase 10.1103/PhysRevLett.96.106802} {\bibfield
  {journal} {\bibinfo  {journal} {Physical Review Letters}\ }\textbf {\bibinfo
  {volume} {96}},\ \bibinfo {pages} {106802} (\bibinfo {year}
  {2006})}\BibitemShut {NoStop}%
\bibitem [{\citenamefont {Qi}\ and\ \citenamefont {Zhang}(2011)}]{Qi2011}%
  \BibitemOpen
  \bibfield  {author} {\bibinfo {author} {\bibfnamefont {X.-L.}\ \bibnamefont
  {Qi}}\ and\ \bibinfo {author} {\bibfnamefont {S.-C.}\ \bibnamefont {Zhang}},\
  }\href {\doibase 10.1103/RevModPhys.83.1057} {\bibfield  {journal} {\bibinfo
  {journal} {Reviews of Modern Physics}\ }\textbf {\bibinfo {volume} {83}},\
  \bibinfo {pages} {1057} (\bibinfo {year} {2011})}\BibitemShut {NoStop}%
\bibitem [{\citenamefont {Khanikaev}\ and\ \citenamefont
  {Shvets}(2017)}]{Khanikaev2017}%
  \BibitemOpen
  \bibfield  {author} {\bibinfo {author} {\bibfnamefont {A.~B.}\ \bibnamefont
  {Khanikaev}}\ and\ \bibinfo {author} {\bibfnamefont {G.}~\bibnamefont
  {Shvets}},\ }\href {\doibase 10.1038/s41566-017-0048-5} {\bibfield  {journal}
  {\bibinfo  {journal} {Nature Photonics}\ }\textbf {\bibinfo {volume} {11}},\
  \bibinfo {pages} {763} (\bibinfo {year} {2017})}\BibitemShut {NoStop}%
\bibitem [{\citenamefont {Wang}\ \emph {et~al.}(2008)\citenamefont {Wang},
  \citenamefont {Chong}, \citenamefont {Joannopoulos},\ and\ \citenamefont
  {Solja{\v{c}}i{\'{c}}}}]{Wang2008}%
  \BibitemOpen
  \bibfield  {author} {\bibinfo {author} {\bibfnamefont {Z.}~\bibnamefont
  {Wang}}, \bibinfo {author} {\bibfnamefont {Y.~D.}\ \bibnamefont {Chong}},
  \bibinfo {author} {\bibfnamefont {J.~D.}\ \bibnamefont {Joannopoulos}}, \
  and\ \bibinfo {author} {\bibfnamefont {M.}~\bibnamefont
  {Solja{\v{c}}i{\'{c}}}},\ }\href {\doibase 10.1103/PhysRevLett.100.013905}
  {\bibfield  {journal} {\bibinfo  {journal} {Physical Review Letters}\
  }\textbf {\bibinfo {volume} {100}},\ \bibinfo {pages} {013905} (\bibinfo
  {year} {2008})}\BibitemShut {NoStop}%
\bibitem [{\citenamefont {Wang}\ \emph {et~al.}(2009)\citenamefont {Wang},
  \citenamefont {Chong}, \citenamefont {Joannopoulos},\ and\ \citenamefont
  {Solja{\v{c}}i{\'{c}}}}]{Wang2009}%
  \BibitemOpen
  \bibfield  {author} {\bibinfo {author} {\bibfnamefont {Z.}~\bibnamefont
  {Wang}}, \bibinfo {author} {\bibfnamefont {Y.}~\bibnamefont {Chong}},
  \bibinfo {author} {\bibfnamefont {J.~D.}\ \bibnamefont {Joannopoulos}}, \
  and\ \bibinfo {author} {\bibfnamefont {M.}~\bibnamefont
  {Solja{\v{c}}i{\'{c}}}},\ }\href {\doibase 10.1038/nature08293} {\bibfield
  {journal} {\bibinfo  {journal} {Nature}\ }\textbf {\bibinfo {volume} {461}},\
  \bibinfo {pages} {772} (\bibinfo {year} {2009})}\BibitemShut {NoStop}%
\bibitem [{\citenamefont {Skirlo}\ \emph {et~al.}(2015)\citenamefont {Skirlo},
  \citenamefont {Lu}, \citenamefont {Igarashi}, \citenamefont {Yan},
  \citenamefont {Joannopoulos},\ and\ \citenamefont
  {Solja{\v{c}}i{\'{c}}}}]{Skirlo2015}%
  \BibitemOpen
  \bibfield  {author} {\bibinfo {author} {\bibfnamefont {S.~A.}\ \bibnamefont
  {Skirlo}}, \bibinfo {author} {\bibfnamefont {L.}~\bibnamefont {Lu}}, \bibinfo
  {author} {\bibfnamefont {Y.}~\bibnamefont {Igarashi}}, \bibinfo {author}
  {\bibfnamefont {Q.}~\bibnamefont {Yan}}, \bibinfo {author} {\bibfnamefont
  {J.}~\bibnamefont {Joannopoulos}}, \ and\ \bibinfo {author} {\bibfnamefont
  {M.}~\bibnamefont {Solja{\v{c}}i{\'{c}}}},\ }\href {\doibase
  10.1103/PhysRevLett.115.253901} {\bibfield  {journal} {\bibinfo  {journal}
  {Physical Review Letters}\ }\textbf {\bibinfo {volume} {115}},\ \bibinfo
  {pages} {253901} (\bibinfo {year} {2015})}\BibitemShut {NoStop}%
\bibitem [{\citenamefont {Longhi}\ \emph {et~al.}(2015)\citenamefont {Longhi},
  \citenamefont {Gatti},\ and\ \citenamefont {Valle}}]{Longhi2015}%
  \BibitemOpen
  \bibfield  {author} {\bibinfo {author} {\bibfnamefont {S.}~\bibnamefont
  {Longhi}}, \bibinfo {author} {\bibfnamefont {D.}~\bibnamefont {Gatti}}, \
  and\ \bibinfo {author} {\bibfnamefont {G.~D.}\ \bibnamefont {Valle}},\ }\href
  {\doibase 10.1038/srep13376} {\bibfield  {journal} {\bibinfo  {journal}
  {Scientific Reports}\ }\textbf {\bibinfo {volume} {5}},\ \bibinfo {pages}
  {13376} (\bibinfo {year} {2015})}\BibitemShut {NoStop}%
\bibitem [{\citenamefont {Lu}\ \emph {et~al.}(2016)\citenamefont {Lu},
  \citenamefont {Joannopoulos},\ and\ \citenamefont
  {Solja{\v{c}}i{\'{c}}}}]{Lu2016}%
  \BibitemOpen
  \bibfield  {author} {\bibinfo {author} {\bibfnamefont {L.}~\bibnamefont
  {Lu}}, \bibinfo {author} {\bibfnamefont {J.~D.}\ \bibnamefont
  {Joannopoulos}}, \ and\ \bibinfo {author} {\bibfnamefont {M.}~\bibnamefont
  {Solja{\v{c}}i{\'{c}}}},\ }\href {\doibase 10.1038/nphys3796} {\bibfield
  {journal} {\bibinfo  {journal} {Nature Physics}\ }\textbf {\bibinfo {volume}
  {12}},\ \bibinfo {pages} {626} (\bibinfo {year} {2016})}\BibitemShut
  {NoStop}%
\bibitem [{\citenamefont {Chen}\ \emph {et~al.}(2017)\citenamefont {Chen},
  \citenamefont {Mei}, \citenamefont {Sun}, \citenamefont {Zhang},
  \citenamefont {Zhao},\ and\ \citenamefont {Wu}}]{Chen2017}%
  \BibitemOpen
  \bibfield  {author} {\bibinfo {author} {\bibfnamefont {Z.-G.}\ \bibnamefont
  {Chen}}, \bibinfo {author} {\bibfnamefont {J.}~\bibnamefont {Mei}}, \bibinfo
  {author} {\bibfnamefont {X.-C.}\ \bibnamefont {Sun}}, \bibinfo {author}
  {\bibfnamefont {X.}~\bibnamefont {Zhang}}, \bibinfo {author} {\bibfnamefont
  {J.}~\bibnamefont {Zhao}}, \ and\ \bibinfo {author} {\bibfnamefont
  {Y.}~\bibnamefont {Wu}},\ }\href {\doibase 10.1103/PhysRevA.95.043827}
  {\bibfield  {journal} {\bibinfo  {journal} {Physical Review A}\ }\textbf
  {\bibinfo {volume} {95}},\ \bibinfo {pages} {043827} (\bibinfo {year}
  {2017})}\BibitemShut {NoStop}%
\bibitem [{\citenamefont {Yang}\ \emph {et~al.}(2019)\citenamefont {Yang},
  \citenamefont {Zhang}, \citenamefont {Wu}, \citenamefont {Dong},
  \citenamefont {Yan},\ and\ \citenamefont {Zhang}}]{Yang2019}%
  \BibitemOpen
  \bibfield  {author} {\bibinfo {author} {\bibfnamefont {B.}~\bibnamefont
  {Yang}}, \bibinfo {author} {\bibfnamefont {H.}~\bibnamefont {Zhang}},
  \bibinfo {author} {\bibfnamefont {T.}~\bibnamefont {Wu}}, \bibinfo {author}
  {\bibfnamefont {R.}~\bibnamefont {Dong}}, \bibinfo {author} {\bibfnamefont
  {X.}~\bibnamefont {Yan}}, \ and\ \bibinfo {author} {\bibfnamefont
  {X.}~\bibnamefont {Zhang}},\ }\href {\doibase 10.1103/PhysRevB.99.045307}
  {\bibfield  {journal} {\bibinfo  {journal} {Physical Review B}\ }\textbf
  {\bibinfo {volume} {99}},\ \bibinfo {pages} {045307} (\bibinfo {year}
  {2019})}\BibitemShut {NoStop}%
\bibitem [{\citenamefont {Hafezi}\ \emph {et~al.}(2011)\citenamefont {Hafezi},
  \citenamefont {Demler}, \citenamefont {Lukin},\ and\ \citenamefont
  {Taylor}}]{Hafezi2011}%
  \BibitemOpen
  \bibfield  {author} {\bibinfo {author} {\bibfnamefont {M.}~\bibnamefont
  {Hafezi}}, \bibinfo {author} {\bibfnamefont {E.~A.}\ \bibnamefont {Demler}},
  \bibinfo {author} {\bibfnamefont {M.~D.}\ \bibnamefont {Lukin}}, \ and\
  \bibinfo {author} {\bibfnamefont {J.~M.}\ \bibnamefont {Taylor}},\ }\href
  {\doibase 10.1038/nphys2063} {\bibfield  {journal} {\bibinfo  {journal}
  {Nature Physics}\ }\textbf {\bibinfo {volume} {7}},\ \bibinfo {pages} {907}
  (\bibinfo {year} {2011})}\BibitemShut {NoStop}%
\bibitem [{\citenamefont {Hafezi}\ \emph {et~al.}(2013)\citenamefont {Hafezi},
  \citenamefont {Mittal}, \citenamefont {Fan}, \citenamefont {Migdall},\ and\
  \citenamefont {Taylor}}]{Hafezi2013}%
  \BibitemOpen
  \bibfield  {author} {\bibinfo {author} {\bibfnamefont {M.}~\bibnamefont
  {Hafezi}}, \bibinfo {author} {\bibfnamefont {S.}~\bibnamefont {Mittal}},
  \bibinfo {author} {\bibfnamefont {J.}~\bibnamefont {Fan}}, \bibinfo {author}
  {\bibfnamefont {A.}~\bibnamefont {Migdall}}, \ and\ \bibinfo {author}
  {\bibfnamefont {J.~M.}\ \bibnamefont {Taylor}},\ }\href {\doibase
  10.1038/nphoton.2013.274} {\bibfield  {journal} {\bibinfo  {journal} {Nature
  Photonics}\ }\textbf {\bibinfo {volume} {7}},\ \bibinfo {pages} {1001}
  (\bibinfo {year} {2013})}\BibitemShut {NoStop}%
\bibitem [{\citenamefont {Gao}\ \emph {et~al.}(2015)\citenamefont {Gao},
  \citenamefont {Lawrence}, \citenamefont {Yang}, \citenamefont {Liu},
  \citenamefont {Fang}, \citenamefont {B{\'{e}}ri}, \citenamefont {Li},\ and\
  \citenamefont {Zhang}}]{Gao2015}%
  \BibitemOpen
  \bibfield  {author} {\bibinfo {author} {\bibfnamefont {W.}~\bibnamefont
  {Gao}}, \bibinfo {author} {\bibfnamefont {M.}~\bibnamefont {Lawrence}},
  \bibinfo {author} {\bibfnamefont {B.}~\bibnamefont {Yang}}, \bibinfo {author}
  {\bibfnamefont {F.}~\bibnamefont {Liu}}, \bibinfo {author} {\bibfnamefont
  {F.}~\bibnamefont {Fang}}, \bibinfo {author} {\bibfnamefont {B.}~\bibnamefont
  {B{\'{e}}ri}}, \bibinfo {author} {\bibfnamefont {J.}~\bibnamefont {Li}}, \
  and\ \bibinfo {author} {\bibfnamefont {S.}~\bibnamefont {Zhang}},\ }\href
  {\doibase 10.1103/PhysRevLett.114.037402} {\bibfield  {journal} {\bibinfo
  {journal} {Physical Review Letters}\ }\textbf {\bibinfo {volume} {114}},\
  \bibinfo {pages} {037402} (\bibinfo {year} {2015})}\BibitemShut {NoStop}%
\bibitem [{\citenamefont {Mittal}\ \emph {et~al.}(2014)\citenamefont {Mittal},
  \citenamefont {Fan}, \citenamefont {Faez}, \citenamefont {Migdall},
  \citenamefont {Taylor},\ and\ \citenamefont {Hafezi}}]{Mittal2014}%
  \BibitemOpen
  \bibfield  {author} {\bibinfo {author} {\bibfnamefont {S.}~\bibnamefont
  {Mittal}}, \bibinfo {author} {\bibfnamefont {J.}~\bibnamefont {Fan}},
  \bibinfo {author} {\bibfnamefont {S.}~\bibnamefont {Faez}}, \bibinfo {author}
  {\bibfnamefont {A.}~\bibnamefont {Migdall}}, \bibinfo {author} {\bibfnamefont
  {J.~M.}\ \bibnamefont {Taylor}}, \ and\ \bibinfo {author} {\bibfnamefont
  {M.}~\bibnamefont {Hafezi}},\ }\href {\doibase
  10.1103/PhysRevLett.113.087403} {\bibfield  {journal} {\bibinfo  {journal}
  {Physical Review Letters}\ }\textbf {\bibinfo {volume} {113}},\ \bibinfo
  {pages} {087403} (\bibinfo {year} {2014})}\BibitemShut {NoStop}%
\bibitem [{\citenamefont {Leykam}\ \emph {et~al.}(2018)\citenamefont {Leykam},
  \citenamefont {Mittal}, \citenamefont {Hafezi},\ and\ \citenamefont
  {Chong}}]{Leykam2018}%
  \BibitemOpen
  \bibfield  {author} {\bibinfo {author} {\bibfnamefont {D.}~\bibnamefont
  {Leykam}}, \bibinfo {author} {\bibfnamefont {S.}~\bibnamefont {Mittal}},
  \bibinfo {author} {\bibfnamefont {M.}~\bibnamefont {Hafezi}}, \ and\ \bibinfo
  {author} {\bibfnamefont {Y.~D.}\ \bibnamefont {Chong}},\ }\href {\doibase
  10.1103/PhysRevLett.121.023901} {\bibfield  {journal} {\bibinfo  {journal}
  {Physical Review Letters}\ }\textbf {\bibinfo {volume} {121}},\ \bibinfo
  {pages} {023901} (\bibinfo {year} {2018})}\BibitemShut {NoStop}%
\bibitem [{\citenamefont {Khanikaev}\ \emph {et~al.}(2013)\citenamefont
  {Khanikaev}, \citenamefont {{Hossein Mousavi}}, \citenamefont {Tse},
  \citenamefont {Kargarian}, \citenamefont {MacDonald},\ and\ \citenamefont
  {Shvets}}]{Khanikaev2012}%
  \BibitemOpen
  \bibfield  {author} {\bibinfo {author} {\bibfnamefont {A.~B.}\ \bibnamefont
  {Khanikaev}}, \bibinfo {author} {\bibfnamefont {S.}~\bibnamefont {{Hossein
  Mousavi}}}, \bibinfo {author} {\bibfnamefont {W.-K.}\ \bibnamefont {Tse}},
  \bibinfo {author} {\bibfnamefont {M.}~\bibnamefont {Kargarian}}, \bibinfo
  {author} {\bibfnamefont {A.~H.}\ \bibnamefont {MacDonald}}, \ and\ \bibinfo
  {author} {\bibfnamefont {G.}~\bibnamefont {Shvets}},\ }\href {\doibase
  10.1038/nmat3520} {\bibfield  {journal} {\bibinfo  {journal} {Nature
  Materials}\ }\textbf {\bibinfo {volume} {12}},\ \bibinfo {pages} {233}
  (\bibinfo {year} {2013})}\BibitemShut {NoStop}%
\bibitem [{\citenamefont {Chen}\ \emph {et~al.}(2014)\citenamefont {Chen},
  \citenamefont {Jiang}, \citenamefont {Chen}, \citenamefont {Zhu},
  \citenamefont {Zhou}, \citenamefont {Dong},\ and\ \citenamefont
  {Chan}}]{Chen2014}%
  \BibitemOpen
  \bibfield  {author} {\bibinfo {author} {\bibfnamefont {W.-J.}\ \bibnamefont
  {Chen}}, \bibinfo {author} {\bibfnamefont {S.-J.}\ \bibnamefont {Jiang}},
  \bibinfo {author} {\bibfnamefont {X.-D.}\ \bibnamefont {Chen}}, \bibinfo
  {author} {\bibfnamefont {B.}~\bibnamefont {Zhu}}, \bibinfo {author}
  {\bibfnamefont {L.}~\bibnamefont {Zhou}}, \bibinfo {author} {\bibfnamefont
  {J.-W.}\ \bibnamefont {Dong}}, \ and\ \bibinfo {author} {\bibfnamefont
  {C.~T.}\ \bibnamefont {Chan}},\ }\href {\doibase 10.1038/ncomms6782}
  {\bibfield  {journal} {\bibinfo  {journal} {Nature Communications}\ }\textbf
  {\bibinfo {volume} {5}},\ \bibinfo {pages} {5782} (\bibinfo {year}
  {2014})}\BibitemShut {NoStop}%
\bibitem [{\citenamefont {Cheng}\ \emph {et~al.}(2016)\citenamefont {Cheng},
  \citenamefont {Jouvaud}, \citenamefont {Ni}, \citenamefont {{Hossein
  Mousavi}}, \citenamefont {Genack}, \citenamefont {Khanikaev}, \citenamefont
  {Mousavi}, \citenamefont {Genack},\ and\ \citenamefont
  {Khanikaev}}]{Cheng2016}%
  \BibitemOpen
  \bibfield  {author} {\bibinfo {author} {\bibfnamefont {X.}~\bibnamefont
  {Cheng}}, \bibinfo {author} {\bibfnamefont {C.}~\bibnamefont {Jouvaud}},
  \bibinfo {author} {\bibfnamefont {X.}~\bibnamefont {Ni}}, \bibinfo {author}
  {\bibfnamefont {S.}~\bibnamefont {{Hossein Mousavi}}}, \bibinfo {author}
  {\bibfnamefont {A.~Z.}\ \bibnamefont {Genack}}, \bibinfo {author}
  {\bibfnamefont {A.~B.}\ \bibnamefont {Khanikaev}}, \bibinfo {author}
  {\bibfnamefont {S.~H.}\ \bibnamefont {Mousavi}}, \bibinfo {author}
  {\bibfnamefont {A.~Z.}\ \bibnamefont {Genack}}, \ and\ \bibinfo {author}
  {\bibfnamefont {A.~B.}\ \bibnamefont {Khanikaev}},\ }\href
  {http://www.ncbi.nlm.nih.gov/pubmed/26901513
  http://www.nature.com/articles/nmat4573 www.nature.com/naturematerials}
  {\bibfield  {journal} {\bibinfo  {journal} {Nature Materials}\ }\textbf
  {\bibinfo {volume} {15}} (\bibinfo {year} {2016})}\BibitemShut {NoStop}%
\bibitem [{\citenamefont {Lai}\ \emph {et~al.}(2016)\citenamefont {Lai},
  \citenamefont {Ma}, \citenamefont {Bo}, \citenamefont {Anlage},\ and\
  \citenamefont {Shvets}}]{Lai2016}%
  \BibitemOpen
  \bibfield  {author} {\bibinfo {author} {\bibfnamefont {K.}~\bibnamefont
  {Lai}}, \bibinfo {author} {\bibfnamefont {T.}~\bibnamefont {Ma}}, \bibinfo
  {author} {\bibfnamefont {X.}~\bibnamefont {Bo}}, \bibinfo {author}
  {\bibfnamefont {S.}~\bibnamefont {Anlage}}, \ and\ \bibinfo {author}
  {\bibfnamefont {G.}~\bibnamefont {Shvets}},\ }\href {\doibase
  10.1038/srep28453} {\bibfield  {journal} {\bibinfo  {journal} {Scientific
  Reports}\ }\textbf {\bibinfo {volume} {6}},\ \bibinfo {pages} {28453}
  (\bibinfo {year} {2016})}\BibitemShut {NoStop}%
\bibitem [{\citenamefont {Slobozhanyuk}\ \emph {et~al.}(2016)\citenamefont
  {Slobozhanyuk}, \citenamefont {Khanikaev}, \citenamefont {Filonov},
  \citenamefont {Smirnova}, \citenamefont {Miroshnichenko},\ and\ \citenamefont
  {Kivshar}}]{Slobozhanyuk2016}%
  \BibitemOpen
  \bibfield  {author} {\bibinfo {author} {\bibfnamefont {A.~P.}\ \bibnamefont
  {Slobozhanyuk}}, \bibinfo {author} {\bibfnamefont {A.~B.}\ \bibnamefont
  {Khanikaev}}, \bibinfo {author} {\bibfnamefont {D.~S.}\ \bibnamefont
  {Filonov}}, \bibinfo {author} {\bibfnamefont {D.~A.}\ \bibnamefont
  {Smirnova}}, \bibinfo {author} {\bibfnamefont {A.~E.}\ \bibnamefont
  {Miroshnichenko}}, \ and\ \bibinfo {author} {\bibfnamefont {Y.~S.}\
  \bibnamefont {Kivshar}},\ }\href {\doibase 10.1038/srep22270} {\bibfield
  {journal} {\bibinfo  {journal} {Scientific Reports}\ }\textbf {\bibinfo
  {volume} {6}},\ \bibinfo {pages} {22270} (\bibinfo {year}
  {2016})}\BibitemShut {NoStop}%
\bibitem [{\citenamefont {Xiao}\ \emph {et~al.}(2016)\citenamefont {Xiao},
  \citenamefont {Lai}, \citenamefont {Yu}, \citenamefont {Ma}, \citenamefont
  {Shvets},\ and\ \citenamefont {Anlage}}]{Xiao2016}%
  \BibitemOpen
  \bibfield  {author} {\bibinfo {author} {\bibfnamefont {B.}~\bibnamefont
  {Xiao}}, \bibinfo {author} {\bibfnamefont {K.}~\bibnamefont {Lai}}, \bibinfo
  {author} {\bibfnamefont {Y.}~\bibnamefont {Yu}}, \bibinfo {author}
  {\bibfnamefont {T.}~\bibnamefont {Ma}}, \bibinfo {author} {\bibfnamefont
  {G.}~\bibnamefont {Shvets}}, \ and\ \bibinfo {author} {\bibfnamefont {S.~M.}\
  \bibnamefont {Anlage}},\ }\href {\doibase 10.1103/PhysRevB.94.195427}
  {\bibfield  {journal} {\bibinfo  {journal} {Physical Review B}\ }\textbf
  {\bibinfo {volume} {94}},\ \bibinfo {pages} {195427} (\bibinfo {year}
  {2016})}\BibitemShut {NoStop}%
\bibitem [{\citenamefont {Yang}\ \emph {et~al.}(2018)\citenamefont {Yang},
  \citenamefont {Xu}, \citenamefont {Xu}, \citenamefont {Wang}, \citenamefont
  {Jiang}, \citenamefont {Hu},\ and\ \citenamefont {Hang}}]{Yang2018}%
  \BibitemOpen
  \bibfield  {author} {\bibinfo {author} {\bibfnamefont {Y.}~\bibnamefont
  {Yang}}, \bibinfo {author} {\bibfnamefont {Y.~F.}\ \bibnamefont {Xu}},
  \bibinfo {author} {\bibfnamefont {T.}~\bibnamefont {Xu}}, \bibinfo {author}
  {\bibfnamefont {H.-X.}\ \bibnamefont {Wang}}, \bibinfo {author}
  {\bibfnamefont {J.-H.}\ \bibnamefont {Jiang}}, \bibinfo {author}
  {\bibfnamefont {X.}~\bibnamefont {Hu}}, \ and\ \bibinfo {author}
  {\bibfnamefont {Z.~H.}\ \bibnamefont {Hang}},\ }\href {\doibase
  10.1103/PhysRevLett.120.217401} {\bibfield  {journal} {\bibinfo  {journal}
  {Physical Review Letters}\ }\textbf {\bibinfo {volume} {120}},\ \bibinfo
  {pages} {217401} (\bibinfo {year} {2018})}\BibitemShut {NoStop}%
\bibitem [{\citenamefont {Chen}\ \emph
  {et~al.}(2018{\natexlab{a}})\citenamefont {Chen}, \citenamefont {Deng},
  \citenamefont {Lu},\ and\ \citenamefont {Dong}}]{Chen2018}%
  \BibitemOpen
  \bibfield  {author} {\bibinfo {author} {\bibfnamefont {X.~D.}\ \bibnamefont
  {Chen}}, \bibinfo {author} {\bibfnamefont {W.~M.}\ \bibnamefont {Deng}},
  \bibinfo {author} {\bibfnamefont {J.~C.}\ \bibnamefont {Lu}}, \ and\ \bibinfo
  {author} {\bibfnamefont {J.~W.}\ \bibnamefont {Dong}},\ }\href {\doibase
  10.1103/PhysRevB.97.184201} {\bibfield  {journal} {\bibinfo  {journal}
  {Physical Review B}\ }\textbf {\bibinfo {volume} {97}},\ \bibinfo {pages}
  {184201} (\bibinfo {year} {2018}{\natexlab{a}})}\BibitemShut {NoStop}%
\bibitem [{\citenamefont {Noh}\ \emph {et~al.}(2018)\citenamefont {Noh},
  \citenamefont {Huang}, \citenamefont {Chen},\ and\ \citenamefont
  {Rechtsman}}]{Noh2018}%
  \BibitemOpen
  \bibfield  {author} {\bibinfo {author} {\bibfnamefont {J.}~\bibnamefont
  {Noh}}, \bibinfo {author} {\bibfnamefont {S.}~\bibnamefont {Huang}}, \bibinfo
  {author} {\bibfnamefont {K.~P.}\ \bibnamefont {Chen}}, \ and\ \bibinfo
  {author} {\bibfnamefont {M.~C.}\ \bibnamefont {Rechtsman}},\ }\href {\doibase
  10.1103/PhysRevLett.120.063902} {\bibfield  {journal} {\bibinfo  {journal}
  {Physical Review Letters}\ }\textbf {\bibinfo {volume} {120}},\ \bibinfo
  {pages} {063902} (\bibinfo {year} {2018})}\BibitemShut {NoStop}%
\bibitem [{\citenamefont {Gao}\ \emph {et~al.}(2017)\citenamefont {Gao},
  \citenamefont {Xue}, \citenamefont {Yang}, \citenamefont {Lai}, \citenamefont
  {Yu}, \citenamefont {Lin}, \citenamefont {Chong}, \citenamefont {Shvets},\
  and\ \citenamefont {Zhang}}]{Gao2017}%
  \BibitemOpen
  \bibfield  {author} {\bibinfo {author} {\bibfnamefont {F.}~\bibnamefont
  {Gao}}, \bibinfo {author} {\bibfnamefont {H.}~\bibnamefont {Xue}}, \bibinfo
  {author} {\bibfnamefont {Z.}~\bibnamefont {Yang}}, \bibinfo {author}
  {\bibfnamefont {K.}~\bibnamefont {Lai}}, \bibinfo {author} {\bibfnamefont
  {Y.}~\bibnamefont {Yu}}, \bibinfo {author} {\bibfnamefont {X.}~\bibnamefont
  {Lin}}, \bibinfo {author} {\bibfnamefont {Y.}~\bibnamefont {Chong}}, \bibinfo
  {author} {\bibfnamefont {G.}~\bibnamefont {Shvets}}, \ and\ \bibinfo {author}
  {\bibfnamefont {B.}~\bibnamefont {Zhang}},\ }\href {\doibase
  10.1038/nphys4304} {\bibfield  {journal} {\bibinfo  {journal} {Nature
  Physics}\ }\textbf {\bibinfo {volume} {14}},\ \bibinfo {pages} {140}
  (\bibinfo {year} {2017})}\BibitemShut {NoStop}%
\bibitem [{\citenamefont {Gladstone}\ \emph {et~al.}(2018)\citenamefont
  {Gladstone}, \citenamefont {Jung},\ and\ \citenamefont
  {Shvets}}]{GladsteinGladstone2018}%
  \BibitemOpen
  \bibfield  {author} {\bibinfo {author} {\bibfnamefont {R.~G.}\ \bibnamefont
  {Gladstone}}, \bibinfo {author} {\bibfnamefont {M.}~\bibnamefont {Jung}}, \
  and\ \bibinfo {author} {\bibfnamefont {G.}~\bibnamefont {Shvets}},\ }\href
  {https://arxiv.org/pdf/1809.02819.pdf http://arxiv.org/abs/1809.02819} {\
  (\bibinfo {year} {2018})},\ \Eprint {http://arxiv.org/abs/1809.02819}
  {arXiv:1809.02819} \BibitemShut {NoStop}%
\bibitem [{\citenamefont {Chen}\ \emph
  {et~al.}(2018{\natexlab{b}})\citenamefont {Chen}, \citenamefont {Shi},
  \citenamefont {Liu}, \citenamefont {Lu}, \citenamefont {Deng}, \citenamefont
  {Dai}, \citenamefont {Cheng},\ and\ \citenamefont {Dong}}]{Chen2018a}%
  \BibitemOpen
  \bibfield  {author} {\bibinfo {author} {\bibfnamefont {X.-D.}\ \bibnamefont
  {Chen}}, \bibinfo {author} {\bibfnamefont {F.-L.}\ \bibnamefont {Shi}},
  \bibinfo {author} {\bibfnamefont {H.}~\bibnamefont {Liu}}, \bibinfo {author}
  {\bibfnamefont {J.-C.}\ \bibnamefont {Lu}}, \bibinfo {author} {\bibfnamefont
  {W.-M.}\ \bibnamefont {Deng}}, \bibinfo {author} {\bibfnamefont {J.-Y.}\
  \bibnamefont {Dai}}, \bibinfo {author} {\bibfnamefont {Q.}~\bibnamefont
  {Cheng}}, \ and\ \bibinfo {author} {\bibfnamefont {J.-W.}\ \bibnamefont
  {Dong}},\ }\href {\doibase 10.1103/PhysRevApplied.10.044002} {\bibfield
  {journal} {\bibinfo  {journal} {Physical Review Applied}\ }\textbf {\bibinfo
  {volume} {10}},\ \bibinfo {pages} {044002} (\bibinfo {year}
  {2018}{\natexlab{b}})}\BibitemShut {NoStop}%
\bibitem [{\citenamefont {Fang}\ \emph
  {et~al.}(2012{\natexlab{a}})\citenamefont {Fang}, \citenamefont {Yu},\ and\
  \citenamefont {Fan}}]{Fang2012a}%
  \BibitemOpen
  \bibfield  {author} {\bibinfo {author} {\bibfnamefont {K.}~\bibnamefont
  {Fang}}, \bibinfo {author} {\bibfnamefont {Z.}~\bibnamefont {Yu}}, \ and\
  \bibinfo {author} {\bibfnamefont {S.}~\bibnamefont {Fan}},\ }\href {\doibase
  10.1103/PhysRevLett.108.153901} {\bibfield  {journal} {\bibinfo  {journal}
  {Physical Review Letters}\ }\textbf {\bibinfo {volume} {108}},\ \bibinfo
  {pages} {153901} (\bibinfo {year} {2012}{\natexlab{a}})}\BibitemShut
  {NoStop}%
\bibitem [{\citenamefont {Fang}\ \emph
  {et~al.}(2012{\natexlab{b}})\citenamefont {Fang}, \citenamefont {Yu},\ and\
  \citenamefont {Fan}}]{Fang2012}%
  \BibitemOpen
  \bibfield  {author} {\bibinfo {author} {\bibfnamefont {K.}~\bibnamefont
  {Fang}}, \bibinfo {author} {\bibfnamefont {Z.}~\bibnamefont {Yu}}, \ and\
  \bibinfo {author} {\bibfnamefont {S.}~\bibnamefont {Fan}},\ }\href
  {http://www.nature.com/doifinder/10.1038/nphoton.2012.236
  www.nature.com/naturephotonics} {\bibfield  {journal} {\bibinfo  {journal}
  {Nature Photonics}\ }\textbf {\bibinfo {volume} {6}},\ \bibinfo {pages} {782}
  (\bibinfo {year} {2012}{\natexlab{b}})}\BibitemShut {NoStop}%
\bibitem [{\citenamefont {Rechtsman}\ \emph
  {et~al.}(2013{\natexlab{a}})\citenamefont {Rechtsman}, \citenamefont
  {Zeuner}, \citenamefont {T{\"{u}}nnermann}, \citenamefont {Nolte},
  \citenamefont {Segev},\ and\ \citenamefont {Szameit}}]{Rechtsman2013}%
  \BibitemOpen
  \bibfield  {author} {\bibinfo {author} {\bibfnamefont {M.~C.}\ \bibnamefont
  {Rechtsman}}, \bibinfo {author} {\bibfnamefont {J.~M.}\ \bibnamefont
  {Zeuner}}, \bibinfo {author} {\bibfnamefont {A.}~\bibnamefont
  {T{\"{u}}nnermann}}, \bibinfo {author} {\bibfnamefont {S.}~\bibnamefont
  {Nolte}}, \bibinfo {author} {\bibfnamefont {M.}~\bibnamefont {Segev}}, \ and\
  \bibinfo {author} {\bibfnamefont {A.}~\bibnamefont {Szameit}},\ }\href
  {\doibase 10.1038/nphoton.2012.302} {\bibfield  {journal} {\bibinfo
  {journal} {Nature Photonics}\ }\textbf {\bibinfo {volume} {7}},\ \bibinfo
  {pages} {153} (\bibinfo {year} {2013}{\natexlab{a}})}\BibitemShut {NoStop}%
\bibitem [{\citenamefont {Rechtsman}\ \emph
  {et~al.}(2013{\natexlab{b}})\citenamefont {Rechtsman}, \citenamefont
  {Zeuner}, \citenamefont {Plotnik}, \citenamefont {Lumer}, \citenamefont
  {Podolsky}, \citenamefont {Dreisow}, \citenamefont {Nolte}, \citenamefont
  {Segev},\ and\ \citenamefont {Szameit}}]{Rechtsman2013a}%
  \BibitemOpen
  \bibfield  {author} {\bibinfo {author} {\bibfnamefont {M.~C.}\ \bibnamefont
  {Rechtsman}}, \bibinfo {author} {\bibfnamefont {J.~M.}\ \bibnamefont
  {Zeuner}}, \bibinfo {author} {\bibfnamefont {Y.}~\bibnamefont {Plotnik}},
  \bibinfo {author} {\bibfnamefont {Y.}~\bibnamefont {Lumer}}, \bibinfo
  {author} {\bibfnamefont {D.}~\bibnamefont {Podolsky}}, \bibinfo {author}
  {\bibfnamefont {F.}~\bibnamefont {Dreisow}}, \bibinfo {author} {\bibfnamefont
  {S.}~\bibnamefont {Nolte}}, \bibinfo {author} {\bibfnamefont
  {M.}~\bibnamefont {Segev}}, \ and\ \bibinfo {author} {\bibfnamefont
  {A.}~\bibnamefont {Szameit}},\ }\href {\doibase 10.1038/nature12066}
  {\bibfield  {journal} {\bibinfo  {journal} {Nature}\ }\textbf {\bibinfo
  {volume} {496}},\ \bibinfo {pages} {196} (\bibinfo {year}
  {2013}{\natexlab{b}})}\BibitemShut {NoStop}%
\bibitem [{\citenamefont {Ma}\ and\ \citenamefont {Shvets}(2017)}]{Ma2017}%
  \BibitemOpen
  \bibfield  {author} {\bibinfo {author} {\bibfnamefont {T.}~\bibnamefont
  {Ma}}\ and\ \bibinfo {author} {\bibfnamefont {G.}~\bibnamefont {Shvets}},\
  }\href {\doibase 10.1103/PhysRevB.95.165102} {\bibfield  {journal} {\bibinfo
  {journal} {Physical Review B}\ }\textbf {\bibinfo {volume} {95}},\ \bibinfo
  {pages} {165102} (\bibinfo {year} {2017})}\BibitemShut {NoStop}%
\bibitem [{\citenamefont {Wang}\ and\ \citenamefont {Fan}(2005)}]{Wang2005}%
  \BibitemOpen
  \bibfield  {author} {\bibinfo {author} {\bibfnamefont {Z.}~\bibnamefont
  {Wang}}\ and\ \bibinfo {author} {\bibfnamefont {S.}~\bibnamefont {Fan}},\
  }\href {\doibase 10.1364/OL.30.001989} {\bibfield  {journal} {\bibinfo
  {journal} {Optics Letters}\ }\textbf {\bibinfo {volume} {30}},\ \bibinfo
  {pages} {1989} (\bibinfo {year} {2005})}\BibitemShut {NoStop}%
\bibitem [{\citenamefont {Qiu}\ \emph {et~al.}(2011)\citenamefont {Qiu},
  \citenamefont {Wang},\ and\ \citenamefont {Solja{\v{c}}i{\'{c}}}}]{Qiu2011}%
  \BibitemOpen
  \bibfield  {author} {\bibinfo {author} {\bibfnamefont {W.}~\bibnamefont
  {Qiu}}, \bibinfo {author} {\bibfnamefont {Z.}~\bibnamefont {Wang}}, \ and\
  \bibinfo {author} {\bibfnamefont {M.}~\bibnamefont {Solja{\v{c}}i{\'{c}}}},\
  }\href {\doibase 10.1364/OE.19.022248} {\bibfield  {journal} {\bibinfo
  {journal} {Optics Express}\ }\textbf {\bibinfo {volume} {19}},\ \bibinfo
  {pages} {22248} (\bibinfo {year} {2011})}\BibitemShut {NoStop}%
\bibitem [{\citenamefont {Zandbergen}\ and\ \citenamefont
  {de~Dood}(2010)}]{Zandbergen2010}%
  \BibitemOpen
  \bibfield  {author} {\bibinfo {author} {\bibfnamefont {S.~R.}\ \bibnamefont
  {Zandbergen}}\ and\ \bibinfo {author} {\bibfnamefont {M.~J.~A.}\ \bibnamefont
  {de~Dood}},\ }\href {\doibase 10.1103/PhysRevLett.104.043903} {\bibfield
  {journal} {\bibinfo  {journal} {Physical Review Letters}\ }\textbf {\bibinfo
  {volume} {104}},\ \bibinfo {pages} {043903} (\bibinfo {year}
  {2010})}\BibitemShut {NoStop}%
\bibitem [{\citenamefont {Bittner}\ \emph {et~al.}(2010)\citenamefont
  {Bittner}, \citenamefont {Dietz}, \citenamefont {Miski-Oglu}, \citenamefont
  {{Oria Iriarte}}, \citenamefont {Richter},\ and\ \citenamefont
  {Sch{\"{a}}fer}}]{Bittner2010}%
  \BibitemOpen
  \bibfield  {author} {\bibinfo {author} {\bibfnamefont {S.}~\bibnamefont
  {Bittner}}, \bibinfo {author} {\bibfnamefont {B.}~\bibnamefont {Dietz}},
  \bibinfo {author} {\bibfnamefont {M.}~\bibnamefont {Miski-Oglu}}, \bibinfo
  {author} {\bibfnamefont {P.}~\bibnamefont {{Oria Iriarte}}}, \bibinfo
  {author} {\bibfnamefont {A.}~\bibnamefont {Richter}}, \ and\ \bibinfo
  {author} {\bibfnamefont {F.}~\bibnamefont {Sch{\"{a}}fer}},\ }\href {\doibase
  10.1103/PhysRevB.82.014301} {\bibfield  {journal} {\bibinfo  {journal}
  {Physical Review B}\ }\textbf {\bibinfo {volume} {82}},\ \bibinfo {pages}
  {014301} (\bibinfo {year} {2010})}\BibitemShut {NoStop}%
\bibitem [{\citenamefont {Kuhl}\ \emph {et~al.}(2010)\citenamefont {Kuhl},
  \citenamefont {Barkhofen}, \citenamefont {Tudorovskiy}, \citenamefont
  {St{\"{o}}ckmann}, \citenamefont {Hossain}, \citenamefont {{de Forges de
  Parny}},\ and\ \citenamefont {Mortessagne}}]{Kuhl2010}%
  \BibitemOpen
  \bibfield  {author} {\bibinfo {author} {\bibfnamefont {U.}~\bibnamefont
  {Kuhl}}, \bibinfo {author} {\bibfnamefont {S.}~\bibnamefont {Barkhofen}},
  \bibinfo {author} {\bibfnamefont {T.}~\bibnamefont {Tudorovskiy}}, \bibinfo
  {author} {\bibfnamefont {H.-J.}\ \bibnamefont {St{\"{o}}ckmann}}, \bibinfo
  {author} {\bibfnamefont {T.}~\bibnamefont {Hossain}}, \bibinfo {author}
  {\bibfnamefont {L.}~\bibnamefont {{de Forges de Parny}}}, \ and\ \bibinfo
  {author} {\bibfnamefont {F.}~\bibnamefont {Mortessagne}},\ }\href {\doibase
  10.1103/PhysRevB.82.094308} {\bibfield  {journal} {\bibinfo  {journal}
  {Physical Review B}\ }\textbf {\bibinfo {volume} {82}},\ \bibinfo {pages}
  {094308} (\bibinfo {year} {2010})}\BibitemShut {NoStop}%
\bibitem [{\citenamefont {Ma}\ \emph {et~al.}(2015)\citenamefont {Ma},
  \citenamefont {Khanikaev}, \citenamefont {Mousavi},\ and\ \citenamefont
  {Shvets}}]{Ma2015}%
  \BibitemOpen
  \bibfield  {author} {\bibinfo {author} {\bibfnamefont {T.}~\bibnamefont
  {Ma}}, \bibinfo {author} {\bibfnamefont {A.~B.}\ \bibnamefont {Khanikaev}},
  \bibinfo {author} {\bibfnamefont {S.~H.}\ \bibnamefont {Mousavi}}, \ and\
  \bibinfo {author} {\bibfnamefont {G.}~\bibnamefont {Shvets}},\ }\href
  {\doibase 10.1103/PhysRevLett.114.127401} {\bibfield  {journal} {\bibinfo
  {journal} {Physical Review Letters}\ }\textbf {\bibinfo {volume} {114}},\
  \bibinfo {pages} {127401} (\bibinfo {year} {2015})}\BibitemShut {NoStop}%
\bibitem [{\citenamefont {Polder}(1948)}]{Polder1949}%
  \BibitemOpen
  \bibfield  {author} {\bibinfo {author} {\bibfnamefont {D.}~\bibnamefont
  {Polder}},\ }\href {\doibase 10.1080/14786444908561215} {\bibfield  {journal}
  {\bibinfo  {journal} {Phys. Rev.}\ }\textbf {\bibinfo {volume} {73}},\
  \bibinfo {pages} {155} (\bibinfo {year} {1948})}\BibitemShut {NoStop}%
\bibitem [{sup()}]{supmat}%
  \BibitemOpen
  \href@noop {} {\ }\BibitemShut {NoStop}%
\bibitem [{\citenamefont {Mung}\ and\ \citenamefont {Chan}(2019)}]{Mung2019}%
  \BibitemOpen
  \bibfield  {author} {\bibinfo {author} {\bibfnamefont {S.~W.}\ \bibnamefont
  {Mung}}\ and\ \bibinfo {author} {\bibfnamefont {W.~S.}\ \bibnamefont
  {Chan}},\ }\href {\doibase 10.1109/MMM.2019.2909518} {\bibfield  {journal}
  {\bibinfo  {journal} {IEEE Microwave Magazine}\ }\textbf {\bibinfo {volume}
  {20}},\ \bibinfo {pages} {55} (\bibinfo {year} {2019})}\BibitemShut {NoStop}%
\bibitem [{\citenamefont {Seif}\ and\ \citenamefont {Hafezi}(2018)}]{Seif2018}%
  \BibitemOpen
  \bibfield  {author} {\bibinfo {author} {\bibfnamefont {A.}~\bibnamefont
  {Seif}}\ and\ \bibinfo {author} {\bibfnamefont {M.}~\bibnamefont {Hafezi}},\
  }\href {\doibase 10.1038/s41566-018-0091-x} {\bibfield  {journal} {\bibinfo
  {journal} {Nature Photonics}\ }\textbf {\bibinfo {volume} {12}},\ \bibinfo
  {pages} {60} (\bibinfo {year} {2018})}\BibitemShut {NoStop}%
\bibitem [{\citenamefont {Fang}\ \emph {et~al.}(2017)\citenamefont {Fang},
  \citenamefont {Luo}, \citenamefont {Metelmann}, \citenamefont {Matheny},
  \citenamefont {Marquardt}, \citenamefont {Clerk},\ and\ \citenamefont
  {Painter}}]{Fang2017}%
  \BibitemOpen
  \bibfield  {author} {\bibinfo {author} {\bibfnamefont {K.}~\bibnamefont
  {Fang}}, \bibinfo {author} {\bibfnamefont {J.}~\bibnamefont {Luo}}, \bibinfo
  {author} {\bibfnamefont {A.}~\bibnamefont {Metelmann}}, \bibinfo {author}
  {\bibfnamefont {M.~H.}\ \bibnamefont {Matheny}}, \bibinfo {author}
  {\bibfnamefont {F.}~\bibnamefont {Marquardt}}, \bibinfo {author}
  {\bibfnamefont {A.~A.}\ \bibnamefont {Clerk}}, \ and\ \bibinfo {author}
  {\bibfnamefont {O.}~\bibnamefont {Painter}},\ }\href {\doibase
  10.1038/nphys4009} {\bibfield  {journal} {\bibinfo  {journal} {Nature
  Physics}\ }\textbf {\bibinfo {volume} {13}},\ \bibinfo {pages} {465}
  (\bibinfo {year} {2017})}\BibitemShut {NoStop}%
\bibitem [{\citenamefont {Shen}\ \emph {et~al.}(2018)\citenamefont {Shen},
  \citenamefont {Zhang}, \citenamefont {Chen}, \citenamefont {Sun},
  \citenamefont {Zou}, \citenamefont {Guo}, \citenamefont {Zou},\ and\
  \citenamefont {Dong}}]{Shen2018}%
  \BibitemOpen
  \bibfield  {author} {\bibinfo {author} {\bibfnamefont {Z.}~\bibnamefont
  {Shen}}, \bibinfo {author} {\bibfnamefont {Y.-L.}\ \bibnamefont {Zhang}},
  \bibinfo {author} {\bibfnamefont {Y.}~\bibnamefont {Chen}}, \bibinfo {author}
  {\bibfnamefont {F.-W.}\ \bibnamefont {Sun}}, \bibinfo {author} {\bibfnamefont
  {X.-B.}\ \bibnamefont {Zou}}, \bibinfo {author} {\bibfnamefont {G.-C.}\
  \bibnamefont {Guo}}, \bibinfo {author} {\bibfnamefont {C.-L.}\ \bibnamefont
  {Zou}}, \ and\ \bibinfo {author} {\bibfnamefont {C.-H.}\ \bibnamefont
  {Dong}},\ }\href {\doibase 10.1038/s41467-018-04187-8} {\bibfield  {journal}
  {\bibinfo  {journal} {Nature Communications}\ }\textbf {\bibinfo {volume}
  {9}},\ \bibinfo {pages} {1797} (\bibinfo {year} {2018})}\BibitemShut
  {NoStop}%
\bibitem [{\citenamefont {Ni}\ \emph {et~al.}(2018)\citenamefont {Ni},
  \citenamefont {Weiner}, \citenamefont {Al{\`{u}}},\ and\ \citenamefont
  {Khanikaev}}]{Ni2018}%
  \BibitemOpen
  \bibfield  {author} {\bibinfo {author} {\bibfnamefont {X.}~\bibnamefont
  {Ni}}, \bibinfo {author} {\bibfnamefont {M.}~\bibnamefont {Weiner}}, \bibinfo
  {author} {\bibfnamefont {A.}~\bibnamefont {Al{\`{u}}}}, \ and\ \bibinfo
  {author} {\bibfnamefont {A.~B.}\ \bibnamefont {Khanikaev}},\ }\href {\doibase
  10.1038/s41563-018-0252-9} {\bibfield  {journal} {\bibinfo  {journal} {Nature
  Materials}\ }\textbf {\bibinfo {volume} {18}} (\bibinfo {year} {2018}),\
  10.1038/s41563-018-0252-9}\BibitemShut {NoStop}%
\bibitem [{\citenamefont {Jia}\ \emph {et~al.}(2019)\citenamefont {Jia},
  \citenamefont {Zhang}, \citenamefont {Gao}, \citenamefont {Guo},
  \citenamefont {Yang}, \citenamefont {Hu}, \citenamefont {Bi}, \citenamefont
  {Xiang}, \citenamefont {Liu},\ and\ \citenamefont {Zhang}}]{Jia2019}%
  \BibitemOpen
  \bibfield  {author} {\bibinfo {author} {\bibfnamefont {H.}~\bibnamefont
  {Jia}}, \bibinfo {author} {\bibfnamefont {R.}~\bibnamefont {Zhang}}, \bibinfo
  {author} {\bibfnamefont {W.}~\bibnamefont {Gao}}, \bibinfo {author}
  {\bibfnamefont {Q.}~\bibnamefont {Guo}}, \bibinfo {author} {\bibfnamefont
  {B.}~\bibnamefont {Yang}}, \bibinfo {author} {\bibfnamefont {J.}~\bibnamefont
  {Hu}}, \bibinfo {author} {\bibfnamefont {Y.}~\bibnamefont {Bi}}, \bibinfo
  {author} {\bibfnamefont {Y.}~\bibnamefont {Xiang}}, \bibinfo {author}
  {\bibfnamefont {C.}~\bibnamefont {Liu}}, \ and\ \bibinfo {author}
  {\bibfnamefont {S.}~\bibnamefont {Zhang}},\ }\href {\doibase
  10.1126/science.aau7707} {\bibfield  {journal} {\bibinfo  {journal}
  {Science}\ }\textbf {\bibinfo {volume} {363}},\ \bibinfo {pages} {148}
  (\bibinfo {year} {2019})}\BibitemShut {NoStop}%
\bibitem [{\citenamefont {Fleury}\ \emph {et~al.}(2014)\citenamefont {Fleury},
  \citenamefont {Sounas}, \citenamefont {Sieck}, \citenamefont {Haberman},\
  and\ \citenamefont {Al{\`{u}}}}]{Fleury2014}%
  \BibitemOpen
  \bibfield  {author} {\bibinfo {author} {\bibfnamefont {R.}~\bibnamefont
  {Fleury}}, \bibinfo {author} {\bibfnamefont {D.~L.}\ \bibnamefont {Sounas}},
  \bibinfo {author} {\bibfnamefont {C.~F.}\ \bibnamefont {Sieck}}, \bibinfo
  {author} {\bibfnamefont {M.~R.}\ \bibnamefont {Haberman}}, \ and\ \bibinfo
  {author} {\bibfnamefont {A.}~\bibnamefont {Al{\`{u}}}},\ }\href {\doibase
  10.1126/science.1246957} {\bibfield  {journal} {\bibinfo  {journal}
  {Science}\ }\textbf {\bibinfo {volume} {343}},\ \bibinfo {pages} {516}
  (\bibinfo {year} {2014})}\BibitemShut {NoStop}%
\bibitem [{\citenamefont {Maayani}\ \emph {et~al.}(2018)\citenamefont
  {Maayani}, \citenamefont {Dahan}, \citenamefont {Kligerman}, \citenamefont
  {Moses}, \citenamefont {Hassan}, \citenamefont {Jing}, \citenamefont {Nori},
  \citenamefont {Christodoulides},\ and\ \citenamefont {Carmon}}]{Maayani2018}%
  \BibitemOpen
  \bibfield  {author} {\bibinfo {author} {\bibfnamefont {S.}~\bibnamefont
  {Maayani}}, \bibinfo {author} {\bibfnamefont {R.}~\bibnamefont {Dahan}},
  \bibinfo {author} {\bibfnamefont {Y.}~\bibnamefont {Kligerman}}, \bibinfo
  {author} {\bibfnamefont {E.}~\bibnamefont {Moses}}, \bibinfo {author}
  {\bibfnamefont {A.~U.}\ \bibnamefont {Hassan}}, \bibinfo {author}
  {\bibfnamefont {H.}~\bibnamefont {Jing}}, \bibinfo {author} {\bibfnamefont
  {F.}~\bibnamefont {Nori}}, \bibinfo {author} {\bibfnamefont {D.~N.}\
  \bibnamefont {Christodoulides}}, \ and\ \bibinfo {author} {\bibfnamefont
  {T.}~\bibnamefont {Carmon}},\ }\href {\doibase 10.1038/s41586-018-0245-5}
  {\bibfield  {journal} {\bibinfo  {journal} {Nature}\ }\textbf {\bibinfo
  {volume} {558}},\ \bibinfo {pages} {569} (\bibinfo {year}
  {2018})}\BibitemShut {NoStop}%
\bibitem [{\citenamefont {Khanikaev}\ \emph {et~al.}(2015)\citenamefont
  {Khanikaev}, \citenamefont {Fleury}, \citenamefont {Mousavi},\ and\
  \citenamefont {Al{\`{u}}}}]{Khanikaev2015}%
  \BibitemOpen
  \bibfield  {author} {\bibinfo {author} {\bibfnamefont {A.~B.}\ \bibnamefont
  {Khanikaev}}, \bibinfo {author} {\bibfnamefont {R.}~\bibnamefont {Fleury}},
  \bibinfo {author} {\bibfnamefont {S.~H.}\ \bibnamefont {Mousavi}}, \ and\
  \bibinfo {author} {\bibfnamefont {A.}~\bibnamefont {Al{\`{u}}}},\ }\href
  {\doibase 10.1038/ncomms9260} {\bibfield  {journal} {\bibinfo  {journal}
  {Nature Communications}\ }\textbf {\bibinfo {volume} {6}} (\bibinfo {year}
  {2015}),\ 10.1038/ncomms9260}\BibitemShut {NoStop}%
\end{thebibliography}

%

\end{document}